\documentstyle[12pt,epsf,bezier,axodraw]{article}
\begin{document}
\setlength{\baselineskip}{0.33 in}
%%%%%%%%%%%%%%%
\catcode`@=11
% Redefine caption to put text and formulas in smaller font
\long\def\@caption#1[#2]#3{\par\addcontentsline{\csname
  ext@#1\endcsname}{#1}{\protect\numberline{\csname
  the#1\endcsname}{\ignorespaces #2}}\begingroup
    \small
    \@parboxrestore
    \@makecaption{\csname fnum@#1\endcsname}{\ignorespaces #3}\par
  \endgroup}
\catcode`@=12
%%%%%%%%%%%%%%%%%%%%%%%%%%%%%%%%%%%%%%%%%%%%%%%%%%%%%%%%%%%%
\newcommand{\newc}{\newcommand}
\newc{\gsim}{\lower.7ex\hbox{$\;\stackrel{\textstyle>}{\sim}\;$}}
\newc{\lsim}{\lower.7ex\hbox{$\;\stackrel{\textstyle<}{\sim}\;$}}
%%%%%%%%%%%%%%%%%% Reference Defs %%%%%%%%%%%%%%%%%

\def\bdm{\begin{equation}}
\def\edm{\end{equation}}
\def\bea{\begin{eqnarray}}
\def\eea{\end{eqnarray}}
\def\ba{\begin{array{}}}
\def\ea{\end{array}}
\def\cc{coupling constant }
\def\msb{$\overline{MS}$ }
\def\vs{v^{\star}}
\def\yts{y_t^{\star}}
\def\mts{m_t^{\star}}
\def\ybs{y_b^{\star}}
\def\mbs{m_b^{\star}}
\def\vs2{v^{\star^2}}
\def\ybs2{y_b^{\star^2}}
\def\mbs2{m_b^{\star^2}}
\def\yts2{y_t^{\star^2}}
\def\mts2{m_t^{\star^2}}
\def\ew{electroweak }
\def\bm{\bar{m}_b}
\def\bmb{\bar{m}_b(\mu)}
\def\bmbt{\bar{m}_b(t)}
\def\bmbm{\bar{m}_b(\bar{m}_b)}
\def\bmbz{\bar{m}_b(M_Z)}
\def\dmm{\frac{\delta m}{m} }
\def\dmu2{\delta\mu^2}
\def\dzvb{\delta\bar{\zeta}_v}
\def\dzvs{\delta\zeta_v^{\star}}
\def\dzv{\delta\zeta_v}
\def\NPB#1#2#3{Nucl. Phys. {\bf B#1} #3 (19#2)}
\def\PLB#1#2#3{Phys. Lett. {\bf B#1} #3 (19#2)}
\def\PRD#1#2#3{Phys. Rev. {\bf D#1} #3 (19#2)}
\def\PRB#1#2#3{Phys. Rev. {\bf B#1} #3 (19#2)}
\def\PRL#1#2#3{Phys. Rev. Lett. {\bf#1} #3 (19#2)}
\def\PRT#1#2#3{Phys. Rep. {\bf#1} #3 (19#2)}
\def\MODA#1#2#3{Mod. Phys. Lett. {\bf A#1} #3 (19#2) }
\def\ZPC#1#2#3{Zeit. f\"ur Physik {\bf C#1} #3 (19#2) }
\def\ZPA#1#2#3{Zeit. f\"ur Physik {\bf A#1} #3 (19#2) }
%%%%%%%%%%%%%%%%%%%%%%%%%%%%%%%%%%%%%%%%%%%%%%%%%%%%%%%%%%%%%
%%%%%%%%%%%%%%%%%%%%%%%%%%%%%%%%%%%%%%%%%%%%%%%%%%%%%%%%
\vsize 8.7in
\def\singlespace{\baselineskip 11.38 pt}
\def\halfagainspace{\baselineskip 17.07 pt}
\def\doublespace{\baselineskip 22.76 pt}
\def\medspace{\baselineskip 17.07 pt}
\font\headings=cmbx10 scaled 1200
\font\title=cmbx10 scaled 1200
\halfagainspace

\begin{titlepage}
\begin{flushright}
{\large

hep-ph/0101291\\
January 20, 2001 \\
}
\end{flushright}
\vskip 2cm
\begin{center}
{\Large {\bf NonQCD contributions to heavy quark masses and sensitivity to Higgs mass}}

\vskip 1cm
{\Large 
R.S. Willey\footnote{E-mail: {\tt willey@pitt.edu}}$^{b}$\\}
\smallskip
$^{b}${\large\it Department of Physics and Astronomy\\
 University of Pittsburgh, Pittsburgh, PA 15260, USA}\\
\end{center}

%%%%%%%%%%%%%%%%%%%%%%%%%%%%%%%%%
\vspace*{.3in}
\begin{abstract}
  We find that if the Higgs mass is close to its present experimental lower      limit (100 GeV),Yukawa interactions in the quark-Higgs sector can make
substantial contribution to the heavy quark MS masses.
\end{abstract}
\end{titlepage}
\setcounter{footnote}{0}
\setcounter{page}{2}
\setcounter{section}{0}
\setcounter{subsection}{0}
\setcounter{subsubsection}{0}
\setcounter{equation}{0}
%%%%%%%%%%%%%%%%%%%%%%%%%%%%%%%%%%%%%
%%%%%%%%%%%%%%%%%%%%%%%%%%%%%%%%%%%%%

\newpage
\section{\bf Introduction}

% body of paper here

   Recent reports from ALEPH,DELPHI, and SLD \cite{ADS} of measurements of the 
\msb mass of the b-quark at the mass scale of the Z mass, combined  
with previous determinations at the scale of the b mass , 
show the running of this mass, and invite
comparison with theory. So far, this has been done entirely in the context
of QCD. In this paper we note that because the t-quark is so heavy its
Yukawa coupling to the scalar sector (including the charged Goldstone 
bosons in Feynman-t'Hooft gauge) is comparable to the strong QCD coupling
constant at the scale of $M_Z$. Here we consider what role these nonQCD
interactions may play in the analysis of these observations.

II One-loop perturbative treatment

 We work in a truncated version of the standard model in which only the 
strong QCD gauge coupling ($g_3$ or $g_s$),the Yukawa couplings of the 
heavy t and b quarks, and the quartic self coupling of the scalar fields
are kept nonzero. The weak gauge couplings ($g_1,g_2$) and all the other
Yukawa couplings are taken to zero.(and $m_b\ll m_t$ is set to zero except
when it is the object being determined). For physical values of these masses
and coupling constants, these approximations are too crude for precision
calculations, but they provide dramatic simplification and hence are very
useful for orientation for more heavy duty calculations, and in some cases,
are sufficiently dominant to provide useful estimates.

  We compute the one-point and two-point functions of the model through 
 one-loop order in \msb renormalization and in on-shell momentum subtraction 
renormalization (OS MOM) in order to determine the counter terms        required for these two different renormalization schemes. The relation between the       perturbative
pole mass and the \msb mass follows from the pole condition
\bdm
     0\;=\;m^{\star}-\bar{m}-\overline{\Sigma}(m^{\star}) \label{defm}
\edm
 
 We use $m$ for generic renormalized mass,$\bar{m}$ for \msb masses, and 
(sometimes) $m^{\star}$ for perturbative pole mass.

(1)\hspace{.2in} t-quark \msb mass

Through two loop order for QCD \cite{2L} and one loop for \ew \cite{bw}\cite{hk}
, the relation for t-quark is 

\bea
 \bar{m}_t(\mu) =  m_t^{\star}\{1-\frac{\alpha_s}{\pi}(\frac{4}{3}+\ln\frac
{\mu^2}{m_t^2}) \hspace{3in} & & \nonumber \\                                    -(\frac{\alpha_s}{\pi})^2((15.37-1.04\;n_f)+                                     (\frac{7}{6}- \frac{185}{24}+\frac{13}{36}n_f)\ln\frac{\mu^2}
{ m_t^2}+(\frac{1}{2}-\frac{11}{8}+\frac{1}{12}n_f)(\ln\frac{\mu^2}{m_t^2})^2)  & & \nonumber  \\
 +\frac{y_t^2}{16\;\pi^2}(-\frac{1}{2}d(r)+\frac{3}{4}\ln{\frac{\mu^2}{m_t^2}}
+\frac{2 N_c}{r^2}(1+\ln{\frac{\mu^2}{m_t^2}})-                                  \frac{\lambda}{16\;\pi^2}(3+3\ln\frac{
\mu^2}{M_h^2})\}  & &   \label{mt}
\eea

Here $N_c =3$ and

\bdm
  r=\frac{M_h}{m_t},  \hspace{1in}  d(r)=-\frac{1}{2}+\int^1_0\;dx(2-x)\ln
(r^2(1-x)+x^2)   \label{defrd}
\edm

 For a perturbative treatment, we input to the right hand side the star 
quantities which are either measurable or expressible as combinations of 
measurable quantities (We consider that the pole mass of a heavy quark
can be determined up to an ambiguity of order $\Lambda_{QCD}$)
\bea
  m_t^{\star}=174 & \hspace{1in} v^{\star}=250 & \hspace{1in} m_b^{\star}=4.9
    \label{input}
\eea
The Higgs mass is put in as one of a series of values ranging from $100$ to
$300$ GeV.
The coupling constants are fixed by the relations
\bdm
   y_i=\sqrt{2}\frac{m_i}{v},\;\;\;i=t,b, \hspace{1in}  
   \lambda=\frac{m_h^2}{2 v^2}  \label{ccvm}
\edm
In particular, $y_t^{\star}=0.9843$ while $y_b^{\star}$ is so small that we  
neglect its
contribution to other quantities.
These relations hold for all orders for bar and star quantities.

   The conventional definition of $the$ \msb mass is $\bar{m}_t(\bar{m}_t)$.
With this definition, the $\ln(\frac{\mu^2}{m_t^2})$'s in (\ref{mt}) are zero,
and (\ref{mt}) reduces to

\bea
 \bar{m}_t=  m_t^{\star}\{1-\frac{\alpha_s}{\pi}(\frac{4}{3}
)-(\frac{\alpha_s}{\pi})^2(15.37-1.04\;n_f)
 & & \nonumber  \\
 +\frac{\yts2}{16\;\pi^2}(-\frac{1}{2}d(r)
+\frac{2 N_c}{r^2})-\frac{\lambda^{\star}}{16\;\pi^2}(3+3\ln\frac{m_t^2}{M_h^2})
 & &   \label{mtb}
\eea

 In 
Table 1. we give values of the nominal \msb top quark mass,$\bar{m_t}$($
\bar{m_t}$), computed with just the QCD contribution ($\alpha_s$) and 
with the nonQCD contribution ($y_t,\lambda$) included. We see sensitivity
to the Higgs mass. For the lightest Higgs considered, $m_h =100$, the 
nonQCD contribution is twice as large as and of opposite sign as the QCD
contribution. For the heaviest Higgs,$m_h=300$, the nonQCD contribution is
only a small correction to the QCD only result.

(2) \hspace{.2in} b-quark \msb mass

  For the b-quark, the flavor independent QCD contribution to the relation
is the same, but the electroweak contribution is different.

\bea
 \bar{m}_b(\mu) =  m_b^{\star}\{1-\frac{\alpha_s}{\pi}(\frac{4}{3}+\ln\frac
{\mu^2}{m_b^2}) \hspace{3in} & & \nonumber \\                                    -(\frac{\alpha_s}{\pi})^2((15.37-1.04\;n_f)+                                     (\frac{7}{6}- \frac{185}{24}+\frac{13}{36}n_f)\ln\frac{\mu^2}
{ m_b^2}+(\frac{1}{2}-\frac{11}{8}+\frac{1}{12}n_f)(\ln\frac{\mu^2}{m_b^2})^2)  & & \nonumber  \\
 +\frac{y_t^2}{16\;\pi^2}(-\frac{5}{8}-\frac{3}{4}\ln(\frac{\mu^2}{m_t^2})
+\frac{2 N_c}{r^2}(1+\ln(\frac{\mu^2}{m_t^2})))-                                 \frac{\lambda}{16\;\pi^2}(3+3\ln\frac{\mu^2}{M_h^2})\} & & \label{mb}
\eea

Now when we set $\mu = m_b$ to get the nominal \msb b-quark mass, the QCD logs
are again zero but in the \ew contribution there remain large logarithms of       $\frac{m_b}{m_t}$. These logs are not the result of running anything from
$\mu=m_t$ to $\mu=m_b$. They arise from two features of the \ew interactions 
which are not present in QCD. First is the existence of large mass
splitting between members of a multiplet, in particular, t and b. The first
$ln\frac{m_t}{m_b}$ in (\ref{mb})comes from the b-quark self-energy diagram
in which a b-quark emits a virtual charged Goldstone boson and propagates as
a t-quark until the boson is reabsorbed. In QCD gluons change color, not 
flavor, and quarks of same flavor, but different color have same mass.
Second is the feature of Spontaneous Symmetry Breaking for mass generation
and the concurrent existence of a scalar field which acquires a vev. When
the scalar field is shifted to get into the proper vacuum, the vev propagates
into the boson 1-,2-,and 3-point functions and the fermion 2-point functions,
and is the source of the second $\ln\frac{m_t}{m_b}$ and the                    $\ln{\frac{M_h}{m_b}}$ in (\ref{mb}). (For more detail, see the appendix).

 For the smaller Higgs masses considered (100,150 GeV), the perturbative mass
shifts with $\mu$ fixed at 4.9              are large enough that the condition $\bar{m}=\bar{m}(\mu=\bar{m})$ is
badly violated, so we let $\mu$ float and solve by iteration until the 
$\mu$ put into (\ref{mb}) is the same as the output $\bar{m}(\mu)$. The 
results are given in Table.2. For Higgs mass of 300 GeV, the \ew contribution
is again a small correction to QCD contribution, but for lighter Higgs, 
including the \ew contribution leads to substantially smaller \msb b-quark
mass than given by QCD contribution alone. For Higgs mass 100 GeV, the 
iteration is unstable and does not converge to any positive value. (The
connection between a ``light'' Higgs and breakdown of \msb perturbation 
theory has been noted before \cite{unstable}).

III Evolution of $\bar{m}_b(\mu)$ from $\mu$ equal $m_b$ to $M_Z$

  From $\mu = 4.9\;\; to \; \; 91.19$ is a large change in scale, leading to
large $\ln[\frac{M_Z^2}{m_b^2}$. 
We therefor differentiate (\ref{mb})  to
obtain the one-loop differential equation for the running of $\bar{m}_b(\mu)$
By keeping just the dominant one-loop QCD and \ew Yukawa terms,  we can obtain  
a simple analytic solution. With $t=\ln(\frac{\mu}{4.9})$
\bdm
 \frac{d}{dt}\bar{m}_b(t) = m^{\star}_b\{\frac{g^2_s(t)}{16 \pi^2}(-8)+
\frac{y_t^2(t)}{16 \pi^2}(\frac{4 N_c}{r^2}\} \equiv -m^{\star}_b A(t)           \label{de}
\edm
\bdm
 \bmbt = m^{\star}_b\exp(-\int_{t_0}^t\;dt^{\prime}A(t^{\prime}))\;
\; + const    \label{des}
\edm
The initial condition
$$ \bar{m}(t_0)=\bar{m}$$
fixes
$$ const = \bar{m}-m^{\star}$$
Then we need the solutions of the one-loop RG equations for the QCD and \ew
Yukawa coupling constants. The one-loop QCD coupling constant is
\bdm
 g_s^2=\frac{g_{s_0}^2}{1+g_{s_0}^2 b_0 t}  \label{gs}
\edm
with
$$
  b_0=\frac{1}{16 \pi^2}(22-\frac{4}{3}n_f)
$$
and $g_{s_0}^2 =2.599$ fixed to make $\alpha_s = .119$ at $\mu=M_Z$.

For the \ew Yukawa coupling constant, we start with the equation relating
$\bar{y}$ and $y^{\star}$ \cite{wb}
\bea
  \bar{y}_t^2  =  y_t^{\star^2}\{ 1-\frac{\alpha_s}{\pi}(\frac{8}{3}-
 2\ln\frac{m_t^2}{\mu^2})  & & \nonumber \\
 & +\frac{y_t^2}{16 \pi^2}(-(N_c+\frac{3}{2})\ln\frac{m_t^2}{\mu^2}+
 \frac{N_c}{2}-d(r)) +\frac{\lambda}{16 \pi^2}(1) & \label{ybs}
\eea
Differentiating, the one-loop RG equation for $\bar{y}_t^2$ is ($y = y_t=
y_{top}$ to be distinguished from $ t=\ln\frac{\mu^2}{4.9}$)
\bdm
 \frac{d}{dt} y^2= \frac{1}{16 \pi^2}((2 N_c+3)(y^{2^2}-12 C_F g_s^2 y^2)
   \label{ydot}
\edm
After substitution of $g_s^2(t)$ (\ref{gs})into (\ref{ydot}) the resulting
differential equation for $y^2(t)$ can be solved.
\bdm
 y^2(t)=y_0^2\{(1+\gamma t)^{\frac{b}{\gamma}}+\frac{a y_0^2}{\gamma -b}(
1+\gamma t)^{\frac{b}{\gamma}}-(1+\gamma t))\}^{-1}  \label{y2sol}
\edm
with
\bea
   a=\frac{9}{16\pi^2}, & \hspace{1in} b=\frac{g_{s_0}^2}{\pi^2} \hspace{1in} & \gamma=b_0 g_{s_0}^2
    \nonumber
\eea
and the initial condition that $\bar{y}_t^2(\mu=m_t)$ is given by (\ref{ybs})
 fixes $y_0^2$.
 With $\bar{g}_s^2(t),\bar{y}_t^2(t)$ determined, the integral (\ref{des})
can be done numerically ($t_Z=\ln\frac{M_Z}{4.9}=2.9237$). 
$$
  \int_{t_0}^{t_Z}\;dt^{\prime}A(t^{\prime})\equiv F
$$
and 
\bdm
  \bar{m}_t(t_Z)= m_b^{\star} e^{-F}  +\bar{m}_b -m_b^{\star}  \label{mbz}
\edm
Results are given in Table.3.

There is no entry in the table for $M_h=100$ GeV because there is no solution
for that $M_h$ in table.2. 
The values of $\bar{m}_b(t_Z)$ obtained for $M_h$ ranging from 150 to 250 GeV
fairly well span the range of experimental values \cite{ADS}
For $M_h=300$ GeV the obtained value is probably high.
 Global fits to \ew precision data \cite{el} disfavor Higgs mass greater 
than 200 GeV. For Higgs mass from 150 to 200 GeV, the values of $\bmbz$
in Table.3 are consistent with the data while the values of $\bmbm$ in Table.2
are not consistent with the generally accepted value (4.1 to 4.3 GeV). This 
leads us to question the provenance of the generally accepted value of
$\bmbm$. 
  It comes from QCD sum rule calculations, Lattice gauge calculations,
and potential models. All done in context of pure QCD. Thus it is necessary
to investigate the contributions of the \ew Yukawa couplings to these 
calculations and see if they might then agree with the smaller values we have     obtained for $\bmbm$ from the transformation (\ref{mb}), in the case of the   smaller Higgs masses.

 The calculations just referred to are 
generally substantial undertakings without the added complication of the 
\ew Yukawa contributions, and we do not propose to repeat and generalize 
any of them in this paper. However, we can identify one dominant term 
 responsible for  most of the \ew  contribution
 seen in Tables 2 and 3, for the lighter Higgs masses.  And to incorporate     that term 
into a simple version of the QCD sum rule analysis is not a major undertaking.

IV Electroweak contribution to QCD sum rule analysis.

  Returning to (\ref{mb}) we see that for light Higgs the term $\frac
{y_t^2}{16 \pi^2}\frac{2 N_c}{r^2}$ is almost the whole \ew contribution.       
This term appears as an \msb counter term for the quark selfenergy function.
Its origin is in the elimination of the tadpoles introduced by the shift of     the Higgs field to do perturbation theory about the stable vacuum. See 
Appendix A and \cite{wb}. In the QCD sum rule calculations, one computes
moments of the dispersion representation of the vacuum polarization.
\bdm
 i\int\;dx e^{iqx}<0|T(j_{\mu}(x)j_{\nu}(0))|0>=(q_{\mu}q_{\nu}-q^2 g_{\mu \nu})(C_I(q^2) + \ldots  \label{cuv}
\edm
\bdm
 C(Q^2)=-\frac{Q^2}{\pi}\int_{4 m^2}^{\infty}\; ds\frac{\Im C(s)}{s(s+Q^2)}
    \label{disp}
\edm
\bdm
 {\cal M}_n = \frac{1}{\pi}\int\; ds\frac{\Im C(s)}{s^{n+1}}  =                  \frac{1}{12 \pi^2 Q_q^2}
\int\;ds\frac{R(s)}{s^{n+1}}   \label{R}
\edm
 
 The QCD Feynman diagrams we include are the basic one-and two-loop vacuum
polarization diagrams (contributions to the moments available in the literature \cite{prpt})  and two diagrams with a single $\dmm$ mass insertion in one or the
other of the quark propagator lines in the first diagram .
The new contribution to $\Im C(s)$  is
\bdm
 \Im\delta C(s) = -\frac{2 N_c}{\pi}\frac{\delta m}{m}\frac{m^4}{s^2}\frac
{1}{\sqrt{1-\frac{4 m^2}{s}}}   \label{dc}
\edm
with
\bdm
 \frac{\delta m}{m} \simeq -\frac{y_t^2}{16 \pi^2}\frac{2 N_c}{r^2}(1-\ln\frac{m_t^2}{m_b^2})
  \label{dmin}
\edm
This gives calculated
\bdm
 {\cal M}_n = {\cal M}^{(0)}_n +{\cal M}^{(1)}_n +\delta{\cal M}^{(1)}_n  
   \label{calc}
\edm

to be matched against the integral of R(s) evaluated from experimental
data on the upsilon resonance plus continuum. 

   In the purely QCD analysis, the convergence of the perturbation
series is acceptable up to fairly high n moments. 

Recently Jamin and Pich \cite{jp} have
done an exhaustive treatment of the purely QCD analysis. They found that
the optimal choice of moments to employ was n from 7 to 15. (Smaller n is 
sensitive to poorly known continuum threshold, and for larger n perturbation
theory gets worse and nonperturbative contributions come into play) Furthermore
the value of the mass obtained is rather stable in this range.
But precisely because the dominant \msb \ew term is so large,the perturbative
treatment fails rapidly as the order n of the moment increases. We give the
results for the moment ${\cal M}_1$ in table.4. For the integral over R in 
(\ref{R})we used PDG data for six upsilon resonances and asymptotic form 
0f R for continuum starting at 11.1 GeV. There is no entry in Table.4
for Higgs mass 100 GeV. For this value of Higgs mass, the perturbative 
calculation gives an unphysical negative value for  ${\cal M}_1$
So again we see the breakdown of the \msb \ew perturbation series 
for this ``light'' Higgs. For the other values of $M_h$ from 150 to 300 GeV,
the values of $\bmbm$ obtained from this crude implementation of the QCD sum    rule,
augmented by \ew contribution, look at least recognizably like the results
in Table.2,obtained from the transformation equation (\ref{mb}). 

V.Conclusions

 Although none of the numerical results presented here are definitive - the 
approximations have been too many and too crude, a consistent pattern has 
emerged, with nonnegligible \ew contributions and values of $\bmbm$ 
substantially smaller than currently accepted, and furthermore, sensitive 
to the mass of the Higgs boson.
  We have started from the position that we have a pretty good knowledge
of the perturbative pole mass of a heavy quark 
           from threshold behavior,potential models,and 
lattice calculations. We have chosen 4.9 GeV (and don't protest if someone
prefers 4.8 or 5.0). We have then used the perturbative definition of the
pole mass and \msb perturbation theory to convert to the nominal $\bm$
mass $\bmbm$. We pointed out some features present in the \ew sector which 
do not appear in QCD. And we found values of $\bm$ which depend on the Higgs
mass, and for range of Higgs masses favored by global fits to precision 
\ew data, are significantly smaller than currently accepted value.  
 At this point in the argument we are in conflict with one or the other of
two accepted positions: $M_h$ should be less than 200 GeV and $\bm$ is 
4.1 to 4.3. We pointed out that this value for $\bm$ is not directly
a measured number, but depends on the analysis used to fit experimental
data. In particular, the analyses which led to this value are all purely
QCD. We then made an (admittedly crude) analysis of what value might 
come out of the QCD sum rule approach if one included the \ew contribution.
The results were gratifyingly similar to those obtained from the conversion
from pole mass. We also carried out the evolution 0f $\bm$ from $\mu=\bm$ 
to $\mu=M_Z$, and found results consistent with the recently reported 
experimental results. There is a possible problem here. Just as in the 
case of $\bmbm$, the analysis presented in the experimental papers was
purely \msb QCD. I have not found a quick way to incorporate the \ew
contribution in this analysis, but I refer to a paper by Rodrigo \cite{rod}
which is aware of the possibility of important \ew Yukawa contributions, but 
claims that they can be suppressed by taking appropriate ratios of ratios of
the experimental data. Then our analysis of the evolution including \ew
contribution would remain consistent with the experimental results.
  Finally, the dominant large \ew contribution we have exploited,$\dzvb $, is 
 specific to the \msb scheme. As explained in the Appendix, in a theory
with a global symmetry and Goldstone theorem,$\dzvs$ may simply be taken to 
be zero in a MOM scheme. There will still be 
a strong  $y_t$ Yukawa coupling connecting the two members of the
 t,b-doublet,which suggests that even with $\dzvs=0$,the
\ew Yukawa interactions may make a contribution to the QCD sum rule analyses.
From the point of view of a MOM scheme, the various $\bm$'s are a nonissue,
and the QCD sum rules become  a more general testing ground for the whole
theory, QCD plus \ew.
In the absence of one dominant contribution, it will be necessary to do
a complete perturbative calculation  
of the  \ew contribution to the QCD moment functions, to see how it works out.

\appendix
\section{details of renormalization schemes}

Since the $2 N_c y^2(\frac{m_t}{M_h^2})$ term is the primary source of our      results which differ from the received wisdom,we explain here in some detail how it
arises and is propagated in the \msb two-point and three-point functions

  The original Lagrangian for the Higgs-Goldstone sector is
\bdm
 {\cal L}=(\partial\Phi^{\dagger}_B)(\partial\Phi_B)-\mu_B^2\Phi^{\dagger}_B
 \Phi_B -\lambda_B(\Phi^{\dagger}_B\Phi_B)^2    \label{A1}
\edm
with

\bdm
   \Phi_B = \left( \begin{array}{rcl}
                          &   \Phi_+^B & \\
                           &  \Phi_0^B &
 \end{array}  \right) =\left( \begin{array}{l}\frac{1}{\sqrt{2}}(\phi_1^B
-i\phi_2^B) \\
\frac{1}{\sqrt{2}}(h^B-\i\phi_3^B) \end{array} \right)
\label{A2}
\edm

  For $\mu_0^2<0$ we are in the broken symmetry phase and $<h_B>$ is not 
zero in the stable vacuum. In order to be perturbing about the true ground
state, rewrite everything in terms of the shifted field which has zero vev
in the true ground state.
\bdm
  <h_B>=V_B,\hspace{1in} h_B = V_B + \hat{h}_B,\hspace{1in} <\hat{h}_B>=0         \label{A3}
\edm
 Renormalize
\bdm
 \begin{array}{c}  \Phi_B = \sqrt{Z_{\phi}} \Phi  \\
                    V_B = \sqrt{Z_{\phi}} V  \\
         \mu^2_B = Z_{\mu^2} \mu^2, \hspace{1in} \lambda_B=Z_{\lambda}\lambda
  \end{array}   \label{A4}
\edm
The quartic scalar vertex renormalization is $Z_4=Z_{\lambda} Z^2_{\phi}$

   The complete all orders vev V is determined  by the 
requirement that the vev of $\hat{h}$ be zero, order by order in the 
loop expansion. Our notation for the order by order expansion of $V$ is
\bdm
 \begin{array}{lcr}  V= \zeta_v v &&  \\
       \zeta_v = 1+\frac{\delta V}{v}& =1+(\zeta_v-1) &= 1+\delta\zeta_v
 \end{array} \label{A5}
\edm
We use the notation $\zeta_v$ rather than $Z_v$, to emphasize that (\ref{A5}) is
not the renormalization of any bare field or bare parameter in the Lagrangian
(\ref{A1}). It is the loop expansion of the vev of the renormalized field (and is 
UV finite in the truncated theory considered here, but both UV divergent 
and gauge variant when the gauge couplings are turned on).

  The field shift (\ref{A3}) generates in the Lagrangian (\ref{A1}) terms linear in
the shifted field.
\bdm
 \begin{array}{c}
   {\cal L}_{linear}= -V(Z_{\phi}Z_{\mu^2}\mu^2+Z_4 \lambda V^2)\hat{h} \\             =-v(\mu^2+\lambda v^2)\hat{h}-\left[((\zeta_v-1)(\mu^2+\lambda v^2))   
      -v((Z_{\phi}Z_{\mu^2}-1)\mu^2 +(Z_4 \zeta_v^2-1)\lambda v^2)\right]       \hat{h}   \end{array}  \label{A6}
\edm
Treated as a perturbing interaction, the first term in (\ref{A6}) will generate 
tree tadpole diagrams. We eliminate them by fixing the tree level vev of
the unshifted field as
\bdm
  v^2 = \frac{-\mu^2}{\lambda}, \hspace{1.5in} (\mu^2 < 0)  \label{A7}
\edm
Then inspection of the terms in the Lagrangian which are quadratic in
the fields after the shift, leads to identification of the masses,
\bdm
    M_{\phi}^2 = 0, \hspace{1.5in}  M_h^2\equiv M^2 =2 \lambda v^2 \label{A8}
\edm
Although this is a relation between quantities identified at tree level,in
our treatment there are no higher order corrections. In an OS MOM scheme this
mass is the physical Higgs mass (modulo problems with unstable particles)
(\cite{unstable}).

   The field shift also generates in (\ref{A1}) terms trilinear in the scalar
fields. There are also trilinear Yukawa couplings between the quarks and the 
scalars, which have been the subject of the bulk of this paper.
These terms generate one-loop tadpole diagrams (fig.1). After the condition
(\ref{A7}) has been applied, (\ref{A6}) is still a term of the Lagrangian linear in the
shifted field, but the coefficients are formally of one-loop order, so the 
tadpole trees generated by it now act as counter terms to the one-loop
tadpole diagrams (fig.1).
\bdm
 {\cal L}^{(1)}_{linear}=v(Z_{\phi}Z_{\mu^2}-Z_4 \zeta_v^2)\frac{M^2}{2}         \hat{h}\equiv v\delta\mu^2\hat{h}   \label{A9}
\edm
The elimination of the one-loop tadpoles fixes this particular combination of
$\delta Z's$ and $\delta\zeta_v$ 
\bdm
  \dmu2 = 3\lambda(A_0+A_M)-2 N_c y^2A_m, \hspace{1in} (m = m_t) \label{A10}
\edm
A is the standard tadpole loop integral, dimensionally regularized.
\bdm
  A_M = i\int_{reg}\;(d\ell)\frac{1}{\ell^2-M^2+\i\delta} =\frac{M^2}{16\pi^2}
  (-\Delta_{\epsilon}+\ln{\frac{M^2}{\mu^2} -1})   \label{A11}
\edm
$$\Delta_{\epsilon}=\frac{2}{\epsilon}-\gamma_E+\ln{4\pi},\hspace{1.5in} 
    \epsilon = 4-d 
$$
In \msb all of the $\delta Z's$ are only of form constant times $\Delta_
{\epsilon}$ and cannot cancel log and constant pieces of combined (\ref{A10}),(\ref{A11}).
Then the condition that the vev of the shifted field vanish in one-loop order
fixes 
\bdm
 \delta\bar{\zeta}_v =\frac{1}{16 \pi^2}\left[3\lambda(1-\ln{\frac{M^2}{\mu^2}})
  -2N_c y^2\frac{m^2}{M^2}(1-\ln{\frac{m_t^2}{\mu^2}})\right] \label{A12}
\edm 
The field shift introduces V, and hence $\dzv$, into the quadratic terms
in (\ref{A1}) as well as generating the linear terms (\ref{A6}).  This is the origin of the
large \ew contribution to the transformation from OS to \msb for $M_h < m_t$.

 In OS MOM renormalization, any of the Z's, as well as $\dzv$. can contribute 
to finite part of (\ref{A10}).  In addition to the condition of the vanishing of 
the vev of the shifted Higgs field, we have to specify a number of 
renormalization conditions equal to the number of bare parameters ($\mu^2_B,
\lambda_B$) and field multiplets ($\Phi_B$) in the Lagrangian. We could 
take those to be the mass shell conditions (pole and residue)
\bdm
 \Sigma_{\phi}(0) = 0,\;\;\;\Sigma_h^{\prime}(M^2)=0, \;\;\;\Sigma_
h(M^2)=0.   \label{A13}
\edm
(For \msb these conditions have zero replaced by finite,i.e. \msb counter       terms remove all  $\Delta_{\epsilon}$'s)
In our notations, the two-point counter terms appear in combinations
\bdm
  \begin{array}{c} D_{\phi}^{-1}(q^2)= q^2-\Sigma_{\phi}^{FD}(q^2)+(Z_{\phi}-1) (q^2)+\dmu2  \\
    D_h^{-1}(q^2)= q^2 -M^2-\Sigma_h^{FD}(q^2)+(Z_{\phi}-1)q^2+\dmu2-(Z_4\zeta_v^2-1)M^2
     \end{array}   \label{A14}
\edm
The same $\dmu2$ (\ref{A10}) satisfies both $<\hat{h}>=0$ and $\Sigma_{\phi}(0)=0$.
This is a manifestation of Goldstone's theorem and is a check on our 
renormalization procedure.  From these conditions one can determine $Z_{\phi},
 Z_{\mu^2}$,and the product $Z_4\zeta_v^2$.
For \msb, $\dzvb$ is separately determined by 
the vanishing vev condition, so all the $\bar{Z}$'s are determined (and it is
verified that $\dzvb$ is finite). For OS MOM scheme another renormalization
condition is required to separate $Z_4$ from $\zeta_v$. For this, we can choose 
$h\phi_+\phi_-$ vertex. This is related to the above two-point functions by 
a Ward Identity.
\bdm
  V \Gamma_{h\phi_+\phi_-}(0,q) = D_{\phi}^{-1}(q^2)-D_h^{-1}(q^2) \label{A15}
\edm
Separate this into tree and one-loop and also write separately integrals from  
Feynman diagrams and counter terms.
\bdm
 \begin{array}{c}
 (v+\delta V)(2\lambda v+\tilde{\Gamma}^{FD} +2\lambda v(Z_4\zeta_v-1))            = \\  q^2-\Sigma_{\phi}^{FD}
+(Z_{\phi}-1)q^2+\dmu2-q^2+M^2+\Sigma_h^{FD}-(Z_{\phi}-1)q^2-\dmu2+(Z_4\zeta_v^2
-1)M^2  \end{array}  \label{A16}
\edm
\bdm
 \begin{array}{c}  v 2\lambda v = M^2  \\
       v \tilde{\Gamma}^{FD}(0,q)=-\Sigma_{\phi}^{FD}(q^2)+\Sigma_h^{FD}(q^2)\\ 
    \delta V 2\lambda v +v 2\lambda v(\delta Z_4 +\dzv)= 0+(\delta Z_4+2\dzv)
 2\lambda v^2     \end{array}  \label{A17}
\edm

 The last line is an identity, The Ward Identity is satisfied for any values
of $\delta Z_4$ and $\dzv$. We complete the the renormalization conditions 
for the OS MOM scheme by the condition
\bdm
  v\Gamma_{h\phi_+\phi_-}(0,q)|_{q^2=M^2}=D_{\phi}^{-1}(M^2)  \label{A18}
\edm
This condition requires $\dzv^{\star}$ to be zero ($V^{\star} = v^{\star}$) and $ Z_4$  is now fixed
by (\ref{A13}) with the Ward Identity then implying (\ref{A18}).

\clearpage

Acknowledgments
  I thank U.Baur for calling the LEP and SLD papers to my attention, I.
Rothstein and A Vainstein for enlightening conversations, and A.Duncan
for advice and encouragement.

\clearpage

% now the references. delete or change fake bibitem. delete next three
%   lines and directly read in your .bbl file if you use bibtex.
%\begin{references}

%\end{references}
\clearpage

% figures follow here
%
% Here is an example of the general form of a figure:
% Fill in the caption in the braces of the \caption{} command. Put the label
% that you will use with \ref{} command in the braces of the \label{} command.
%
% \begin{figure}
% \caption{}
% \label{}
% \end{figure}

 \begin{figure}
 \caption{One-loop tadpole diagrams with counter term}
 \label{fig1}
 \epsffile{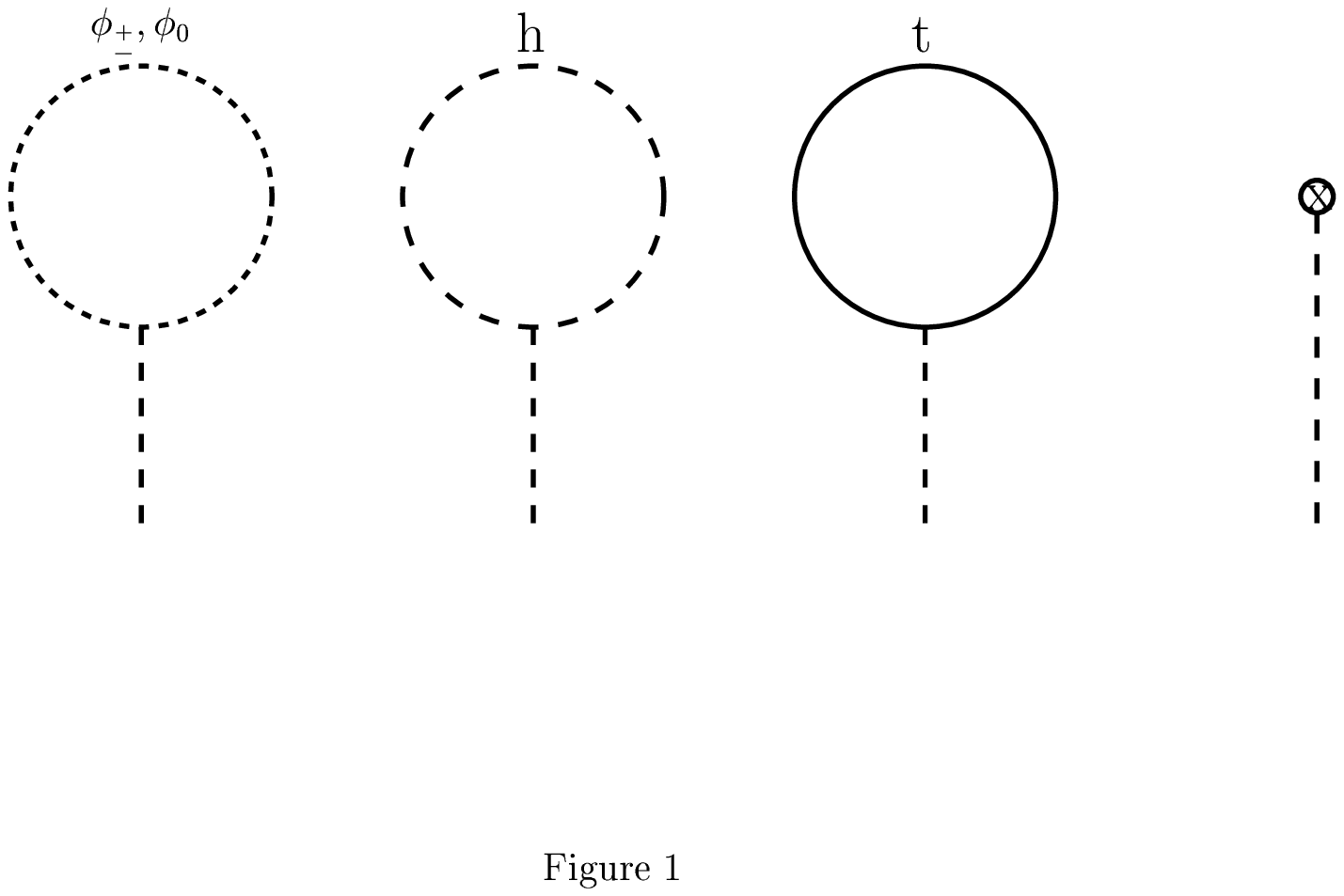}
 \end{figure}

% \begin{figure}
% \caption{extrapolation for \Sp (top) and \Ss (bottom)}
% \label{fig2}
% \epsffile{cmpqplot.eps}
% \end{figure}

% tables follow here
%
% Here is an example of the general form of a table:
% Fill in the caption in the braces of the \caption{} command. Put the label
% that you will use with \ref{} command in the braces of the \label{} command.
% Insert the column specifiers (l, r, c, d, etc.) in the empty braces of the
% \begin{tabular}{} command.
%
 \begin{table}
\caption{t-quark \msb mass $\bar{m}_t(\bar{m}_t$)} \hspace{1.5in}
 \label{Table.1}
 \begin{tabular}{cccc}
 $M_h$ & QCD(1-L) & QCD(2-L) & QCD+EW  \\   \hline
  100 & 166 & 164 & 185   \\  
  150 & 166   & 164     & 174  \\
  200 & 166   & 164     & 170  \\
  250 & 166   & 164     & 168.6  \\
  300 & 166   & 164     & 168
 
\end{tabular}
 \end{table}

 \begin{table}
\caption{b-quark \msb mass $\bar{m}_b(\bar{m}_b$)}  \hspace{1.5in}
 \label{Table.2}
 \begin{tabular}{rccl}
 $M_h$ & QCD(1-L) & QCD(2-L) & QCD+EW  \\   \hline
  100 & 4.46 & 4.19 & --   \\  
  150 & 4.46   & 4.19     & 2.52  \\
  200 & 4.46   & 4.19     & 3.60  \\
  250 & 4.46   & 4.19     & 4.11  \\
  300 & 4.46   & 4.19     & 4.47
 
\end{tabular}
 \end{table}

\begin{table}
\caption{b-quark \msb mass at $M_Z$, $\bar{m}_b(M_Z)$}  \hspace{1.5in}
\label{Table.3}
 \begin{tabular}{ccccc}
 $M_h$ & $\bar{m}$ & $t_0$ & F  &  $\bar{m}_b(M_Z)$  \\   \hline
  100 & --    &           &        &   \\  
  150 &  2.52 &     -.665  & .0528  &  2.27   \\
  200 &  3.60 &     -.308  & .1679  &  2.84    \\
  250 &  4.11  &    -.196  & .2133 &  3.17    \\
  300 &  4.47  &    -.0918 & .2343  & 3.45
 
\end{tabular}
 \end{table}

 \begin{table}
\caption{b-quark \msb mass $\bar{m}_b$ from sum rule}  \hspace{1.5in}
 \label{Table.4}
 \begin{tabular}{ccc}
 $M_h$ & QCD & QCD+EW  \\   \hline
  100 & 4.46 &  --   \\  
  150 & 4,46  & 2.86  \\
  200 & 4.46  & 3.63  \\
  250 & 4.46  & 3.98  \\
  300 & 4.46  & 4.19
 
\end{tabular}
 \end{table}

\end{document}